\documentclass[10pt]{article}
\usepackage{soul}
\usepackage{tikz}
\usepackage{amsmath}
\usepackage{amsfonts}
\newcommand{\kinectTM}{Kinect\texttrademark}
\newcommand{\kinectTMS}{\kinectTM~}
\newcommand{\Prob}{\mathbb{P}}
\newcommand{\Sf}{\textit{S}}

\newcommand{\Ob}{\textit{O}}
\newcommand{\D}{\mathcal{D}}
\newcommand{\Lg}{\mathcal{L}}
\newcommand{\Z}{\textit{Z}}
\newcommand{\J}{J}
\newcommand{\avg}[1]{\langle #1 \rangle}

\usepackage{xcolor}

\usepackage{epstopdf}

\title{Path-integral representation of diluted pedestrian dynamics} 

\author{
Alessandro Corbetta\\
\small Department of Applied Physics,\\
\small Eindhoven University of Technology,\\
\small P.O.Box 513, 5600MB, Eindhoven, the Netherlands, \\
\small \texttt{a.corbetta@tue.nl} \\
\\
Federico Toschi\\
\small Department of Applied Physics and\\
\small Department of Mathematics and Computer Science,\\
\small Eindhoven University of Technology,\\
\small P.O.Box 513, 5600MB, Eindhoven, the Netherlands, and \\
\small CNR-IAC, Rome, Italy,\\
\small \texttt{f.toschi@tue.nl}
}

\begin{document}

\maketitle

\begin{abstract} 
We frame the issue of pedestrian dynamics modeling in terms of path-integrals, a formalism originally introduced in quantum mechanics to account for the behavior of quantum particles, later extended to  quantum field theories and to statistical physics. Path-integration enables a  trajectory-centric representation of the pedestrian motion, directly providing the probability of observing a given trajectory. This appears as the most natural language to describe the statistical properties of pedestrian dynamics in generic settings.
 In a given venue,  individual trajectories can belong to many possible usage patterns and, within each of them, they can display wide variability.

We provide first a primer on path-integration, and we introduce and discuss the path-integral functional probability measure for pedestrian dynamics in the diluted limit. As an illustrative example, we connect the path-integral description to a Langevin model that we developed previously for a particular crowd flow condition (the flow in a narrow corridor). Building on our previous real-life measurements, we provide a quantitatively correct path-integral representation for this condition. Finally, we show how the path-integral formalism can be used to evaluate the probability of rare-events (in the case of the corridor, U-turns).
\end{abstract}

\section{Introduction}\label{ESRAvSMAP:sec:introduction}

Modeling the dynamics of walking pedestrians is a longstanding issue, characterized by a high societal relevance and by fascinating scientific challenges. 
How do people walk and interact in crowds? What influences the motion of single individuals? What is the role of environmental conditions on their dynamics? Which design features can optimize crowd evacuation efficiency? These are among the many -and mostly open- fundamental and engineering questions sustaining an ever growing interest in  pedestrian dynamics modeling (for an overview on the field, we refer to general reviews~\cite{cristiani2014multiscale,helbing2001traffic}, while for model calibration see, e.g.,~\cite{hoogendoorn2007microscopic,seer2014kinects,corbetta2015parameter}). 

While some stunning emergent feature of pedestrian flows, such as the spontaneous  formation of lanes in counter-flow scenarios~\cite{helbing2001self}, or the alternating behavior across bottlenecks~\cite{seyfried2009new},  have been successfully modeled in qualitative terms, a systematic  quantitative comprehension of the crowd motion  allowing for reliable  predictions is still far, and subject of ongoing research. Generic crowd flow settings usually come as combinations of large individual variabilities and the simultaneous presence of several, often location-specific, usage patterns: a daunting challenge for modeling.

\begin{figure}[t!h!]
\center{
  \includegraphics[width=.48\textwidth]{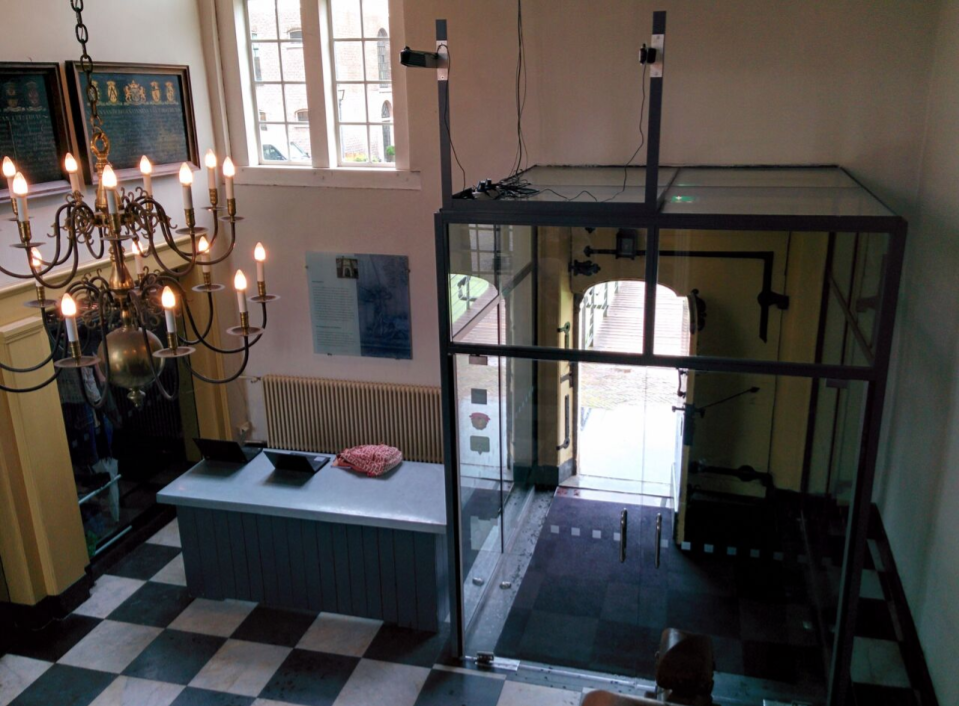}
  \includegraphics[width=.48\textwidth,trim=0.5cm 3.2cm 0cm 2.8cm,clip=true]{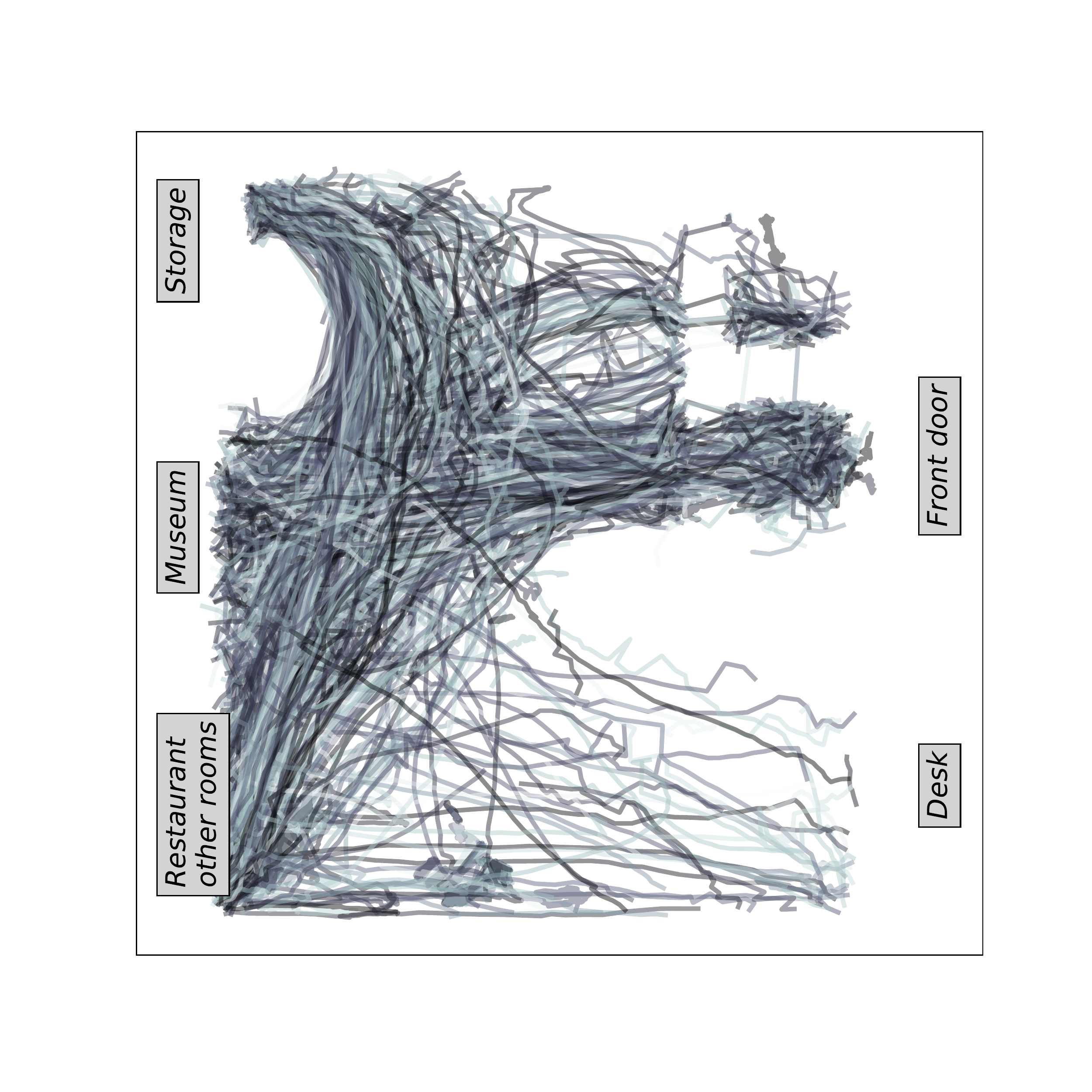}}
\caption{(left) Entrance of the Pesthuis, Naturalis Museum (Leiden, NL) - picture of the front entrance door. The entrance, beside allowing visitors into the museum, gives access to the restaurant and a storage area (accessible only to employees). Since 2016, we installed two overhead tracking devices (Microsoft \kinectTMS~\cite{Kinect}, in conjunction to proprietary tracking technology - for details we refer the reader to our previous publications~\cite{corbetta2016fluctuations,corbetta2014TRP}) to automatically acquire the trajectories of walking pedestrians. We report the portrait of trajectories acquired over one week of measurement in the (right) panel. Despite the variability in the trajectories, few dominant behavior are clearly observable (cf. Figure~\ref{fig:naturalis-cluster}). 
\label{fig:naturalis-setup}}
\end{figure}

\begin{figure}[t!h!]
  \includegraphics[width=.48\textwidth]{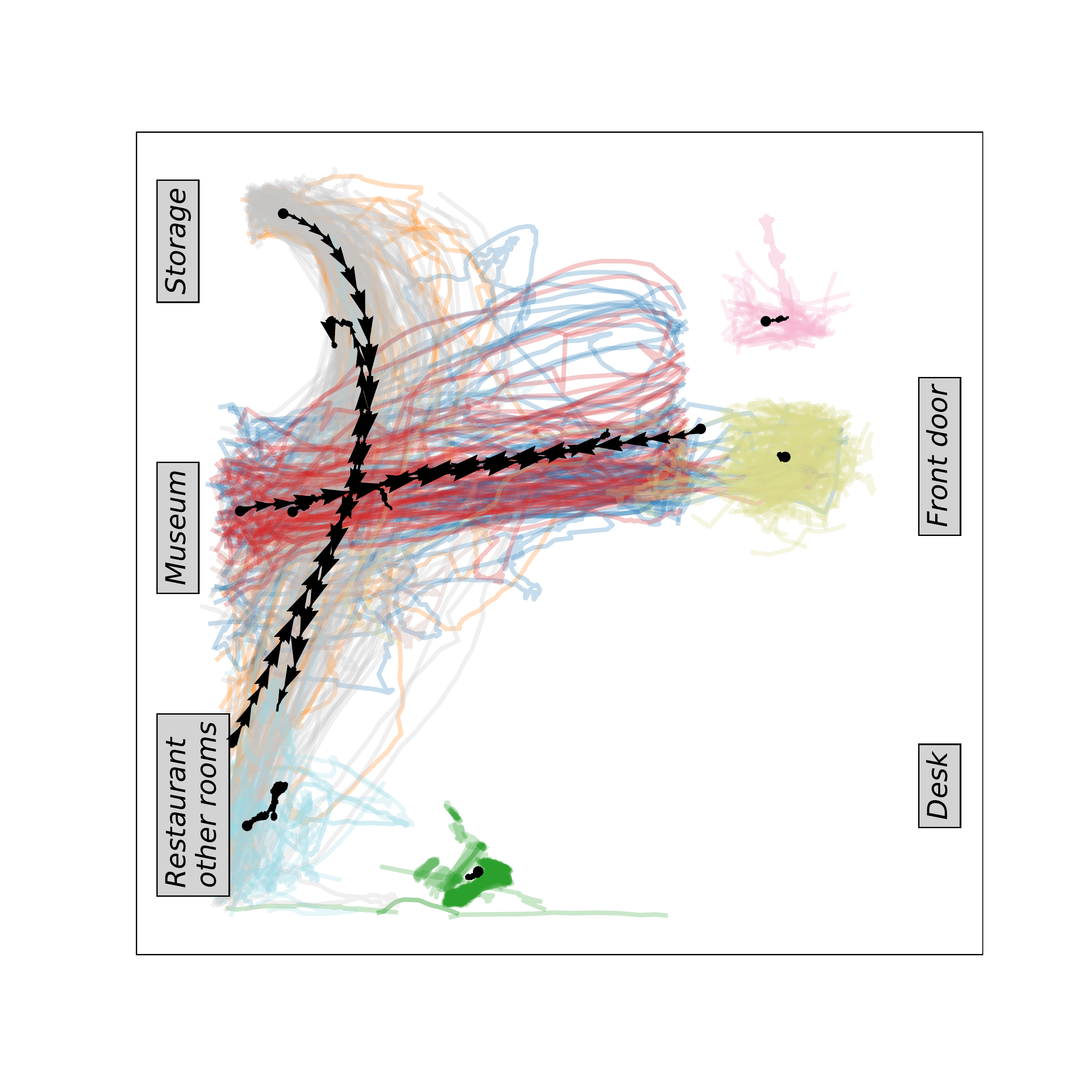}
  \includegraphics[width=.48\textwidth]{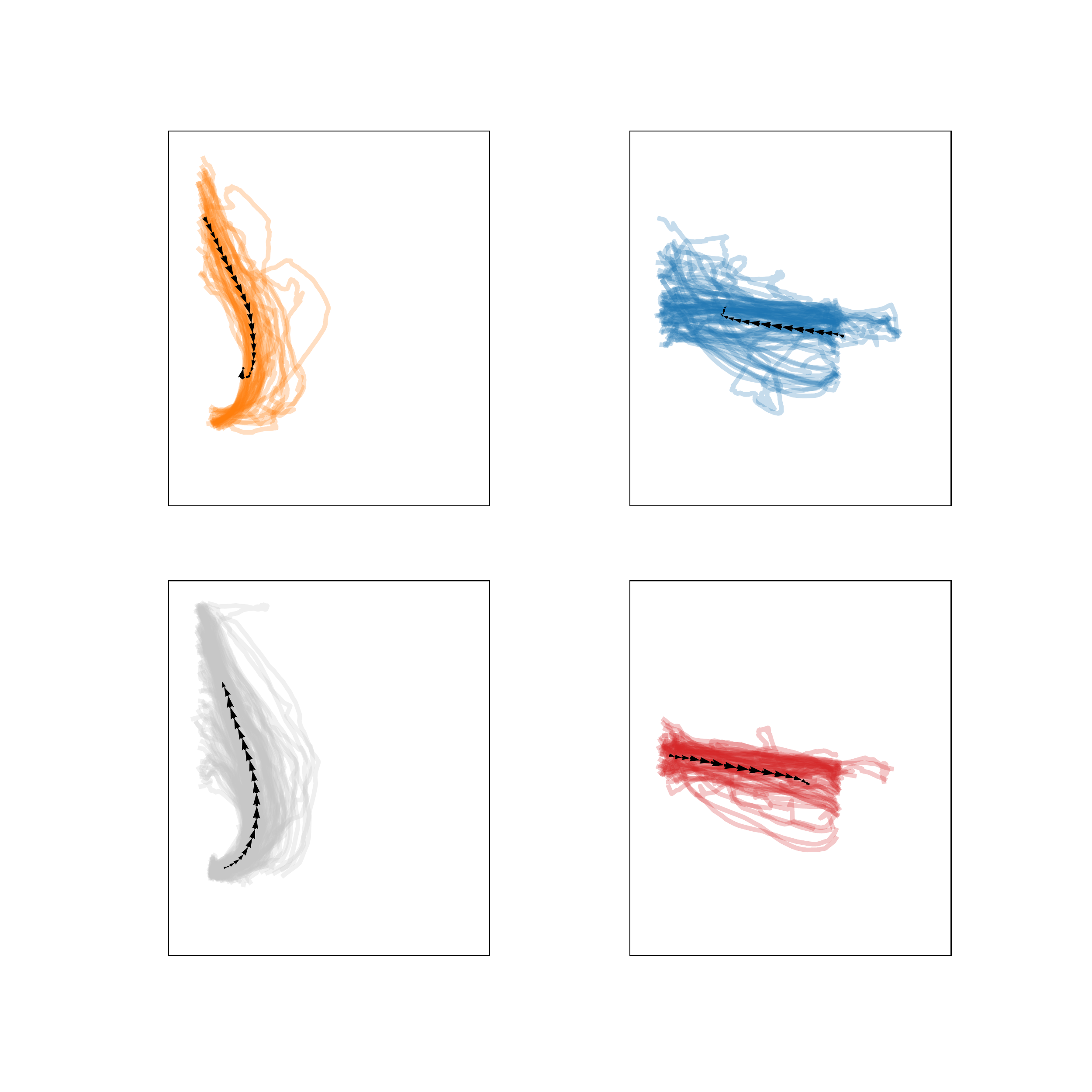}
\caption{(left) Trajectories clusters in Figure~\ref{fig:naturalis-setup} are colorized and outliers are removed. Clusters have been isolated through the DBSCAN clustering algorithm~\cite{ester1996density}, while, for simplicity, the clustering metric between trajectories is the euclidean distance in the four dimensional space $(x(t_1),y(t_1),x(t_2),y(t_2))$. In other words trajectories belong to the same cluster if their ending points are close. (right) The four  clusters to which the largest number of trajectories belong. These encompass employees walking between storage area and restaurant and visitors walking between the entrance and the museum. \label{fig:naturalis-cluster}}
\end{figure}

Individual trajectories, e.g. in a wide public space, can exhibit randomness originating from variability in individual behaviors. First, there is a variability in destination and in purpose, for which the individual paths target one specific destination amongst the many possibly available. Second, there is a variability in  the reaction to external stimuli: a point of interest can attract just few individuals; peer pedestrians can attract  and/or repel others and so on. In Figure~\ref{fig:naturalis-setup}, we report a collection of pedestrian trajectories acquired by us in the public atrium  of a natural science museum (Naturalis Biodiversity Center, Leiden, NL; more details about the measurements in the figure caption). The atrium is a connection zone, and it is crossed by visitors directed to different parts of the museum. In agreement with intuition, we observe  ample variability in trajectories and  a relatively  wide portion of the floor area remain used. However, not all the  trajectories that are physically possible are observed and, in particular, not all trajectories appear to be equally likely. Beside few rare trajectories, clearly largely dissimilar from all others (filtered out in Figure~\ref{fig:naturalis-cluster}(left)),  four main usage patterns, represented by four trajectory clusters, emerged (cf. Figure~\ref{fig:naturalis-cluster}(right)). 

In this chapter we introduce a mathematical representation for pedestrian motion rooted around individual trajectories,   possibly the most intuitive and natural representation of pedestrian motion.  This representation aims at key questions as:  which paths (and under which conditions) are most likely pursued? How wide are the characteristic fluctuations within these paths? And also, which rare events are to be expected? How frequently (rare) dangerous events occur? 
This representation is based on the  known path-integral formalism from quantum mechanics, which assigns to each trajectory  the probability that it is  observed. In the context of pedestrian dynamics this representation and its relevance are still unexplored. So far, microscopic models, based on the analogy between pedestrians and particles (e.g.~\cite{helbing2001traffic,hoogendoorn2007microscopic}), have been a preferred (yet not exclusive, e.g.~\cite{cristiani2014multiscale,schadschneider2001cellular}) choice  to model pedestrian behavior and its variabilities.  Microscopic models prescribe a dynamics via the time-evolution of individual positions and trajectories are recovered by time-integration (cf. primer in Section~\ref{sect:micro-model}).

The path-integral representation associates to a walking trajectory, $\gamma$, the probability of its occurrence  (the trajectory $\gamma$ is thus the time mapping $\gamma: t\mapsto \vec{x}(t) = (x(t),y(t))$, for $t\in[t_i,t_f]$, where $\vec{x}(t)$ is the pedestrian position at time $t$). By formally indicating with $\D\gamma$ the (infinite) measure over all possible trajectories (which we formally build in Section~\ref{sect:derivation}), this probability, with density $\rho[\gamma]$, reads
\begin{equation}\label{eq:path-integral-intro}
d\Prob[\gamma] = \rho[\gamma]\,\D\gamma = \frac{1}{M} e^{-\Sf[\gamma]}\,\D\gamma,
\end{equation}
for a given action functional $\Sf[\gamma]$ and an appropriate normalization constant, that we generically indicate with $M$. The action functional $\Sf[\gamma]$ incorporates a comprehensive knowledge of the  properties of the motion and allows relevant insights. The general notion in field theory is that the knowledge of the action functional $\Sf[\gamma]$ represents the theory itself (see e.g.~\cite{zinn1996quantum}).  
According to Eq.~\eqref{eq:path-integral-intro}, trajectories in the neighborhood of  local minimizers of $\Sf[\gamma]$ are most likely observed as they maximize the observation probability $\Prob$. 
Moreover, action minimizing trajectories, say $\gamma_m$, identify ``average motions'' around which the majority of the  observed trajectories concentrate. For a minimizer $\gamma_m$, the  necessary condition $\delta \Sf[\gamma_m] =0$ must hold, i.e. the variation of the action must vanish for $\gamma_m$. 

Equation~\eqref{eq:path-integral-intro} enables the evaluation of the moments of all possible observable quantities, $\Ob[\gamma]$, built out of the trajectories. The expected value of $\Ob[\gamma]$ (in symbols: $\avg{\Ob[\gamma]}$), for instance, satisfies:
\begin{equation}\label{eq:O-example-avg}
\avg{\Ob[\gamma]} = \int e^{-\Sf[\gamma]} \Ob[\gamma]\, \D \gamma .
\end{equation}
More general momenta can be defined through the moment-generating functional, $Z[J]$, which satisfies:
\begin{equation}\label{eq:Z-intro}
\Z[\J] = \int e^{-\Sf[\gamma]+J\cdot \gamma}\D\gamma,
\end{equation}
where the scalar product is defined as
\begin{equation}
J\cdot \gamma = \int J(t) \cdot \gamma(t)\, dt.
\end{equation}
Thus, through $\J$, the average trajectory is written as 
\begin{equation}\avg{\gamma(t)} = \frac{\delta \log \Z[\J]}{\delta \J(t)}
\end{equation}
or the two-points correlation as
\begin{equation}
\avg{\gamma(t_a)\gamma(t_b)} = \frac{\delta^2 \log \Z[\J]}{\delta \J(t_a)\delta \J(t_b)},
\end{equation}
and analogously, through $N$-th order functional differentiation, we can obtain the $N$-point correlation function. For the details on functional differentiation operators we refer, e.g., to~\cite{zinn1996quantum}.

The content of this chapter is structured as follows: in Section~\ref{sect:micro-model} we give a primer on microscopic modeling of pedestrian dynamics in terms of Langevin equations. In Section~\ref{sect:derivation} we derive formally the action functional $\Sf$ for Langevin dynamics; in Section~\ref{sect:example}, building on our previous works, we derive a quantitative expression for the path-integral in the case of a narrow corridor. The chapter will be concluded with the discussion section~\ref{sect:discussion}, about the use  of path-integrals as natural language and theory to describe the dynamics of pedestrians in most general conditions.

\section{Microscopic modeling of pedestrian dynamics in the diluted limit}\label{sect:micro-model}
Microscopic models  and, specifically, Langevin-like equations~\cite{helbing2001traffic,Lutz}, have been often  employed to describe the pedestrian motion since the beginning of  the its systematic study  by the physics community~\cite{helbing1995social}. Langevin-like equations treat  pedestrians as Newton-like particles whose acceleration is proportional to the superposition of deterministic  forces and random solicitations. These forces are not the outcome of physical interactions, rather they model \textit{social} interactions~\cite{helbing1995social}. We remark that, beside the pedestrian dynamics case, Langevin equations have ubiquitous use in the modeling of physical systems exhibiting random dynamics, and they are employed to model  both passive~\cite{lemons1997paul} and active ``self-propelled''~\cite{Lutz} matter.  Notably, action functionals, and therefore path-integrals, can be written in explicit form for dynamics expressed via  Langevin equations, as we show in Section~\ref{sect:derivation}.

In this chapter we focus on pedestrian dynamics in the diluted limit, i.e. when extremely low pedestrian density and interactions among individuals are absent or negligible. This is the case when people walk alone or when their  distance with the closest individual in the surrounding crowd is sufficiently large. 
In this condition, we model the motion of an individual as the Langevin dynamics:
\begin{equation}
\left\{
\begin{array}{l}
\dot{\vec{v}}(t) = - \frac{\partial \phi}{\partial \vec{v}}(\vec{v}) - \frac{\partial V}{\partial \vec{x}}(\vec{x}) + \sigma\dot {\vec{\eta}} \\
\dot{\vec{x}}(t) = \vec{v}(t).
\end{array}
\right.\label{eq:langevin-gen}
\end{equation}
Here, $\vec{v}(t)$ is the walking velocity of a pedestrian in $\vec{x}$ and $\sigma\dot{\vec{ \eta}}$ is a stochastic forcing term encompassing variabilities and random external influences. 
For simplicity we assume that the stochastic term $\sigma\dot \eta$ is a white uncorrelated Gaussian noise.
Two additional terms, decoupled for simplicity, influence the dynamics: the velocity potential, $\phi(\vec{v})$, and the position confinement potential, $V(\vec{x})$. 

The velocity potential models the fact that pedestrians are self-propelled agents: they can convert internal energy into motion around preferred average velocities. Let $\vec{v}_m$ be a preferred velocity, then $\vec{v}_m$ is a local minimum of $\phi$ (i.e., in the vicinity of $\vec{v}_m$, and for some $\alpha >0$,  the first order approximation  $\phi(\vec{v}_m + \delta \vec{v}) - \phi(\vec{v}_m) = \alpha | \vec{v}_m + \delta \vec{v}|^2$ holds) and, for small noise, the dynamics remains confined around $\vec{v}_m$. In the last part of the chapter we will investigate the case $\phi(u) = \alpha (u-u_m)^2(u+u_m)^2$ for the dynamics in a narrow corridor (here $u$ is the component of $\vec{v}$ parallel to the walking direction). This assumes that pedestrians have preferred average velocities $\pm u_m$ corresponding to the two opposite  walking directions.  In~\cite{corbetta2016fluctuations}, we showed that this model allows to capture quantitatively the statistics of the motion, including fluctuations, as well as, the occurrence of rare events. For a corridor, rare events are U-turns, i.e. velocity inversions $\vec{v}\rightarrow -\vec{v}$. We will discuss this result in view of path-integrals in Section~\ref{sect:pathint-long}.  $V(\vec{x})$, instead, aims at modeling the surrounding environment and therefore it can include repulsion of obstacles, attraction of points of interests~\cite{kwak2013collective} and so on.

\section{Path-integral representation for  pedestrian dynamics}\label{sect:derivation}
In this section we  derive the expression of the action $S[\gamma]$ and of the probability of observing the pedestrian trajectory $\gamma$ (cf. Eq.~\eqref{eq:path-integral-intro}) for a Langevin-like dynamics~\eqref{eq:langevin-gen}. For simplicity, we  operate in the scalar case (i.e. one spatial dimension), as the generalization to the vector case involves  only  small technical complications. Our derivation follows~\cite{chow2015path}, to which we refer for further details.

We partition the interval $[t_i,t_f]$  in $N$ equal segments $[t_j,t_{j+1}]$, with $j=0,1,\ldots,N-1$, of length $\Delta t$. Therefore, it holds 
\begin{equation*}
t_j = j\Delta t.
\end{equation*}
We  express the probability $\rho[\gamma]\,\D\gamma$ as the joint probability of observing the configuration 
\begin{equation}\label{eq:joint-probability}
\Prob(\gamma(t_0),\, \gamma(t_1),\,\ldots\,,\, \gamma(t_N) )
\end{equation}
in the formal limit $\Delta t \rightarrow 0$, i.e. $N\rightarrow \infty$. For a more compact notation, we will  write $\gamma_j$ to indicate $\gamma(t_j)$. 

Because of  our choice of a $\delta$-correlated white noise, the joint probability~\eqref{eq:joint-probability} factorizes with Markov property  as
\begin{equation}\label{eq:markov-factorization}
\Prob(\gamma_0,\, \gamma_1,\,\ldots\,,\, \gamma_N) = \Prob(\gamma_0) \cdot \prod_{j=0}^{N-1} \Prob(\gamma_{j+1}  | \gamma_{j} ).
\end{equation}
Following  the It\^o calculus convention (see, e.g.,~\cite{klebaner2012introduction}) the position $x_j$ and velocity $v_j$ for $\gamma_j$ read
\begin{equation}
\left\{
\begin{array}{l}
v_{j+1} = v_{j} - l_j\Delta t + \sigma \sqrt{\Delta t} \xi_j   \\
x_{j+1} = x_{j} + v_{j}\Delta t,
\end{array}
\right.\label{eq:langevin-discr}
\end{equation}
where we set 
\begin{equation}
l_j = \frac{\partial \phi}{\partial v}(v_n) +  \frac{\partial V}{\partial x}(x_n)
\end{equation}
and $\xi_j$ follows a  centered Gaussian distribution with unit variance. 
We can write the factors $\Prob(\gamma_{j+1}  | \gamma_{j} )$ in explicit form as 
\begin{align}
\Prob(\gamma_{j+1}  | \gamma_{j} ) &= \int d\Prob(\xi_j) \delta ( v_{j+1} - v_{j} + l_j\Delta t + \sigma \sqrt{\Delta t} \xi_j )\label{eq:cond-1}\\
&= \frac{1}{M} \iint d\xi_j\, d\omega\, e^{-i\omega (v_{j+1} - v_{j} + l_j\Delta t + \sigma \sqrt{\Delta t} \xi_j ) - \xi_j^2/2 }\label{eq:cond-2}, 
\end{align}
where in~\eqref{eq:cond-1} $\delta$ denotes the Dirac delta function, which, in~\eqref{eq:cond-2}, is cast in Fourier representation (via the relation $\delta(s) = 1/M \int d\omega e^{i\omega s}$). Note that the function to be integrated in~\eqref{eq:cond-2} satisfies 
\begin{align}
&e^{i\omega (v_{j+1} - v_{j} + l_j\Delta t + \sigma \sqrt{\Delta t} \xi ) - \frac{\xi^2}{2}} =\\
&\qquad= e^{-\frac{1}{2}(\xi - i \omega \sigma \sqrt{\Delta t})^2 -\frac{1}{2}\omega^2\sigma^2\Delta t + i\omega(v_{j+1} - v_{j} + l_j\Delta t )} \\
&\qquad= e^{-\frac{1}{2}(\xi - i \omega \sigma \sqrt{\Delta t})^2  -\frac{1}{2 \sigma^2 \Delta t}(\omega + i (v_{j+1} - v_{j} + l_j\Delta t) )^2  - \frac{1}{2 \sigma^2 \Delta t}(v_{j+1} - v_{j} + l_j\Delta t)^2}\\
&\qquad=e^{-\frac{1}{2}(\xi + i \omega \sigma \sqrt{\Delta t})^2}e^{-\frac{1}{2 \sigma^2 \Delta t}(\omega - i (v_{j+1} - v_{j} + l_j\Delta t) )^2}e^{-\frac{1}{2 \sigma^2 \Delta t}(v_{j+1} - v_{j} + l_j\Delta t)^2} \label{eq:cond-7}.
\end{align}
The first two factors provide normalization constants for~\eqref{eq:cond-7}, therefore we obtain
\begin{equation}
\Prob(\gamma_{j+1}  | \gamma_{j} ) = \frac{1}{M}e^{-\frac{\Delta t}{2 \sigma^2 }(\frac{v_{j+1} - v_{j}}{\Delta t} + l_j)^2}.
\end{equation} 
The product in~\eqref{eq:markov-factorization} thus reads
\begin{align}
\prod_{j=0}^{N-1} \Prob(\gamma_{j+1}  | \gamma_{j} ) &= \frac{1}{M} \prod_{j=0}^{N-1} e^{-\frac{\Delta t}{2 \sigma^2 }(\frac{v_{j+1} - v_{j}}{\Delta t} + l_j)^2}\\
&= \frac{1}{M}  e^{-\frac{\Delta t}{2 \sigma^2 }(\sum_{j=0}^{N-1}\frac{v_{j+1} - v_{j}}{\Delta t} + l_j)^2}\label{eq:discrete-path-integral},
\end{align}
which in the formal limit $\Delta t \rightarrow 0$ yields
\begin{equation}
\frac{1}{M}  e^{-\frac{\Delta t}{2 \sigma^2 }(\sum_{j=0}^{N-1}\frac{v_{j+1} - v_{j}}{\Delta t} + l_j)^2} \rightarrow \frac{1}{M}e^{-\frac{1}{2\sigma^2} \int_{t_i}^{t_f} dt (\dot{v} + \partial_v\phi + \partial_x V)^2 }.
\end{equation}
The probability density of observing a trajectory $\gamma$, $\rho[\gamma]$, is hence
\begin{equation}
\rho[\gamma] = \frac{1}{M}e^{-\frac{1}{2\sigma^2} \int_{t_i}^{t_f} dt (\dot{v} + \frac{\partial\phi}{\partial v} + \frac{\partial V}{\partial x})^2 },
\end{equation}
where 
\begin{equation}
 S[\gamma] =  \frac{1}{2\sigma^2} \int_{t_i}^{t_f} dt \left(\dot{v} + \frac{\partial\phi}{\partial v} + \frac{\partial V}{\partial x }\right)^2
\end{equation}
is the action (cf.~\eqref{eq:path-integral-intro}). In this context, $S[\gamma]$ is also referred to as Onsager-Machlup functional (cf. e.g.~\cite{durr1978onsager}).

Hence, up to a normalization constant, the functional differential $\D \gamma$ is to be understood in the limit sense
\begin{equation}
\D \gamma = \frac{1}{M}\lim_{\Delta t \rightarrow 0} \prod_{j=0}^{N-1}dv_j.
\end{equation}
 For a Langevin dynamics, the trajectories for which $S[\gamma]$ is stationary (i.e. those for which the variation $\delta S[\gamma]$ vanishes) and, in particular, minimum, identify the trajectories observed with highest likelihood. It is well known that these solve the Euler-Lagrange equation 
\begin{equation}
 \frac{\partial \Lg}{\partial  x} -  \frac{d}{dt}\frac{\partial \Lg}{\partial \dot x} + \frac{d^2}{dt^2}\frac{\partial \Lg}{\partial \ddot x} = 0,
\end{equation}
for the Lagrangian function
\begin{equation}\label{eq:lagrangian-def}
\Lg(\ddot x, \dot x, x) = \frac{1}{2\sigma^2}\left(\ddot{x} + \frac{\partial\phi}{\partial \dot{x}} + \frac{\partial V}{\partial x}\right)^2.
\end{equation}

\section{Langevin dynamics in a narrow corridor}\label{sect:example}
In this section we consider the dynamics of pedestrian in a very simple scenario: a narrow corridor. This setting resembles an almost one dimensional geometry as there exists one preferred ``longitudinal'' direction of motion, i.e. along the corridor span. For this case we verified experimentally~\cite{corbetta2016fluctuations} that the pedestrian motion follows quantitatively a Langevin-like dynamics for a proper choice of the potentials $\phi$ and $V$.  This means that the dynamics exhibits, in quantitative terms, the same statistical features of a Langevin motion (including the probability distribution functions of the  walking position and velocity, and the related autocorrelation functions). We  first discuss  the mathematical  model and introduce its experimental verification. Then,  we  obtain its associated path-integral representation, which is therefore also experimentally correct, and we employ it  to derive estimates for the probability of occurrence of the rare events of the dynamics. 
The content of this section relies on and expands our previous works~\cite{corbetta2016fluctuations,corbetta2014TRP,corbetta2016continuous,corbettaTGF15} to which we refer for further details, in particular all those connected to the measurements.  

\begin{table}
\begin{center}
\caption{Parameters used in the model. $\alpha$: modulating factor of the double-well potential force $f$ governing the longitudinal motion (cf.~\eqref{m2});  $\beta$: stiffness coefficient of the transversal linear Langevin dynamics; $\gamma$:  friction coefficient of the transversal linear Langevin dynamics; $\sigma$: white noise intensity in longitudinal and traversal direction; $u_m$: desired mean walking speed.} \label{tabparam}
\begin{tabular}{|l r l  | l r l| }
\hline
$\alpha$ &  $0.0625$ & m$^{-2}$s & $\sigma$ &  $0.16$ & ms$^{-3/2}$ \\
$\beta$ & $1.63$ & s/m$^{-2}$ &  &  & \\
$\gamma$ &  $0.207$ & s$^{-1}$ & $u_m$ &   $1.0$ & ms$^{-1}$ \\
\hline
\end{tabular}
\end{center}
\end{table}

\begin{figure}[t!h!]
\begin{center}
\includegraphics[width=.37\textwidth,trim=0.5cm .5cm 0cm 5.8cm,clip=true]{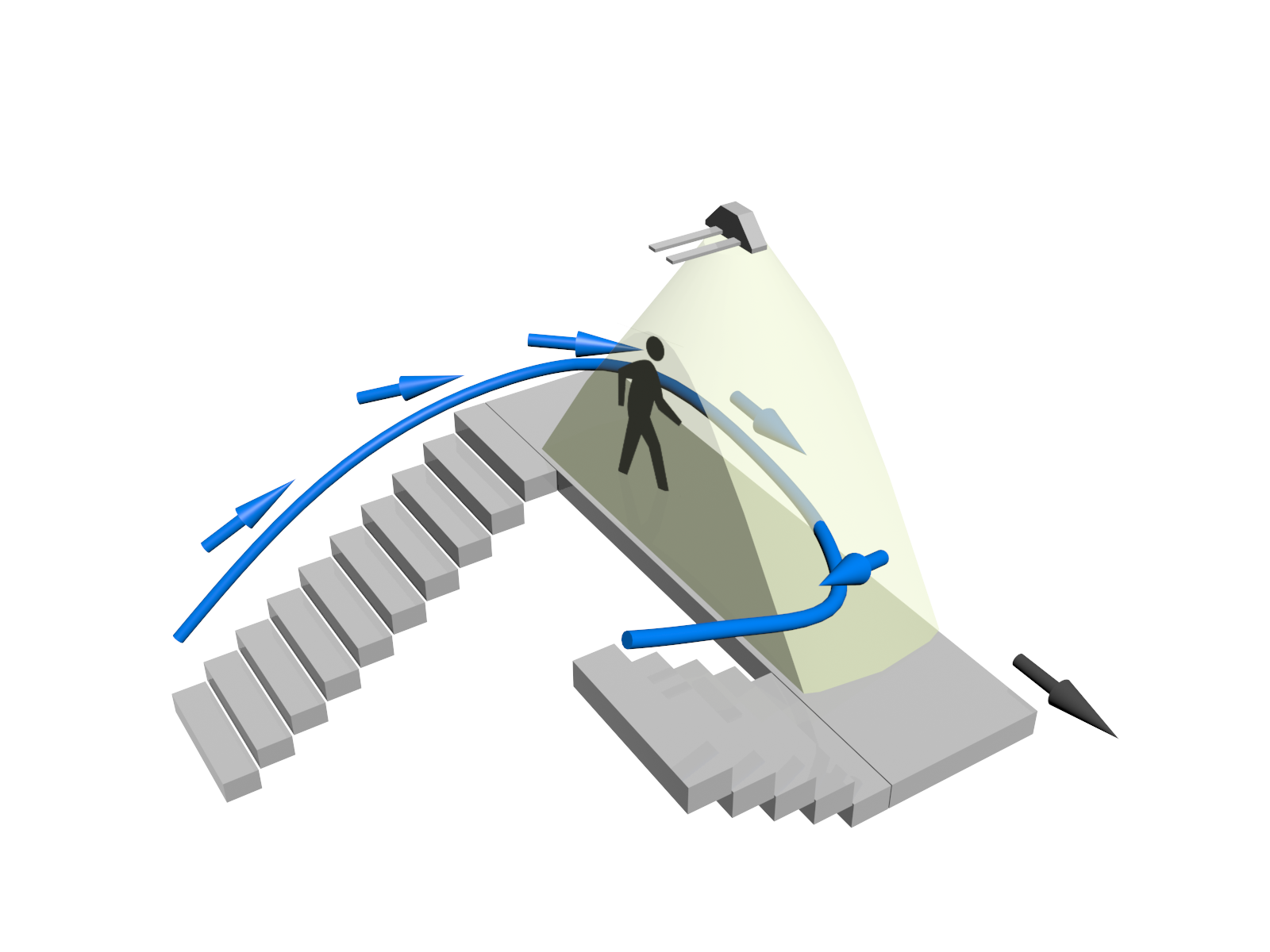}
\includegraphics[width=.60\textwidth,trim=0.5cm 8.0cm 2.5cm 8.8cm,clip=true]{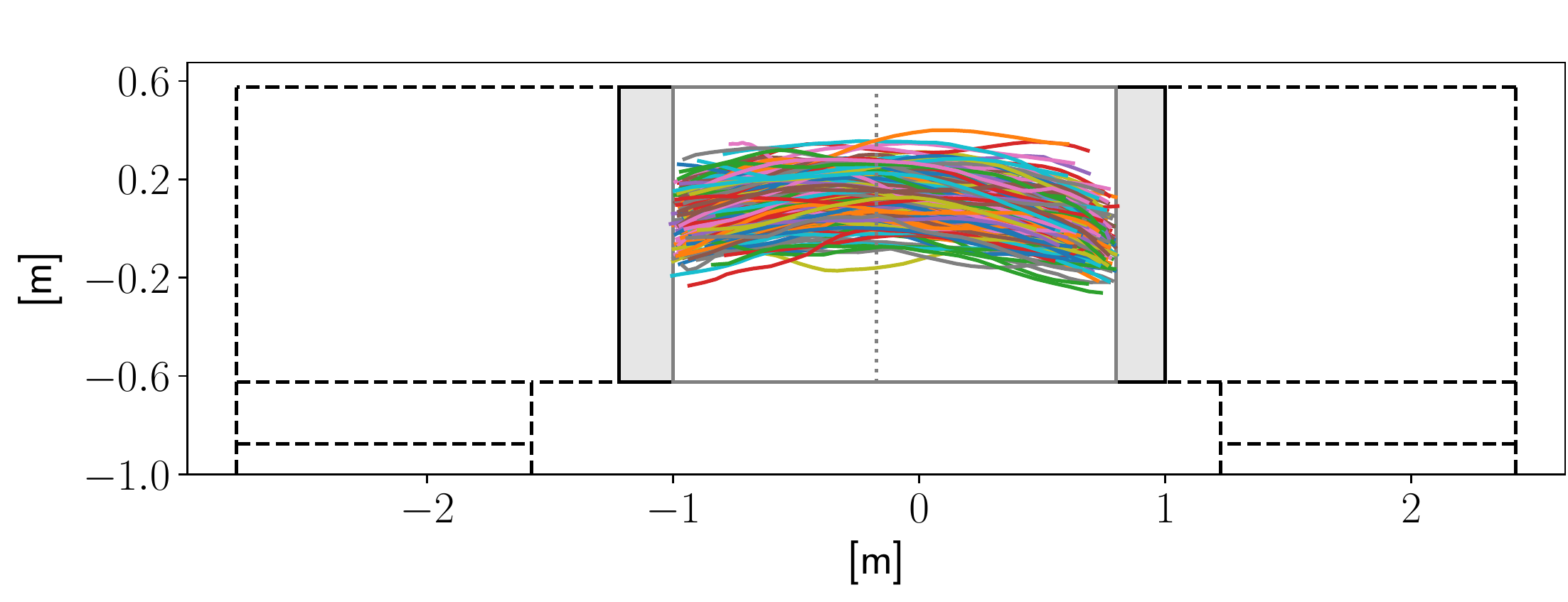}
\end{center}
\caption{(a) 3D sketch of the corridor in the Metaforum building of TU/e where the experimental observations were made, with a representation of the  view cone of the  \kinectTM. An ideal sample trajectory (analogous to panel b) from the left-end to the right-end of the facility is reported. The whole area is surrounded by walls, here removed for readability. (Figure reproduced from~\cite{CDA10}) (b) Examples of real-life trajectories recorded, involving pedestrians crossing the facility  (Figure reproduced from~\cite{datasetped}).  
\label{fig:corridor}}
\end{figure}

\begin{figure}
\centering
\begin{tikzpicture} 
\node at (0,0){
     \includegraphics[width=5.8cm]{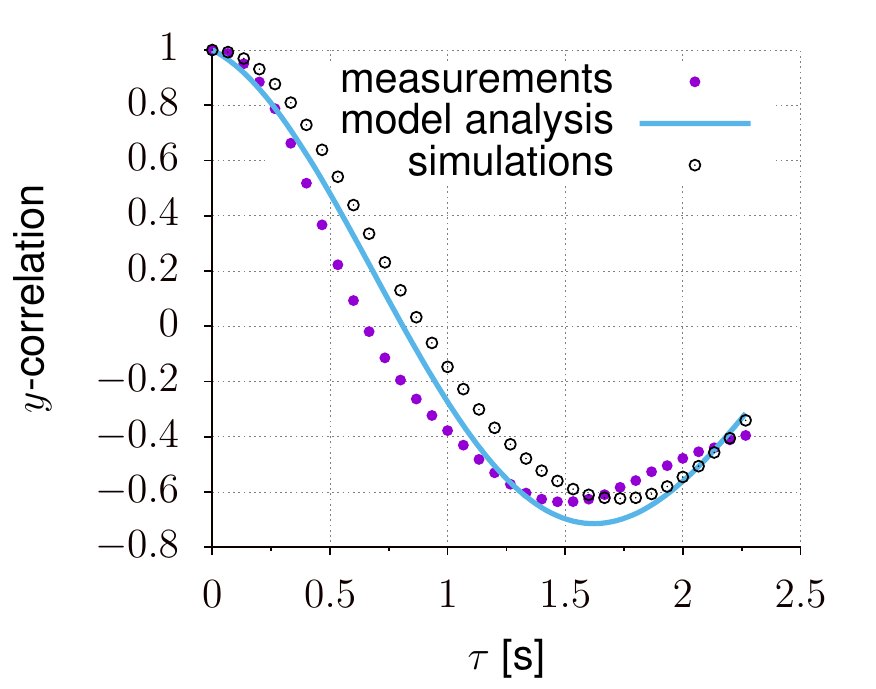}
     \includegraphics[width=5.8cm]{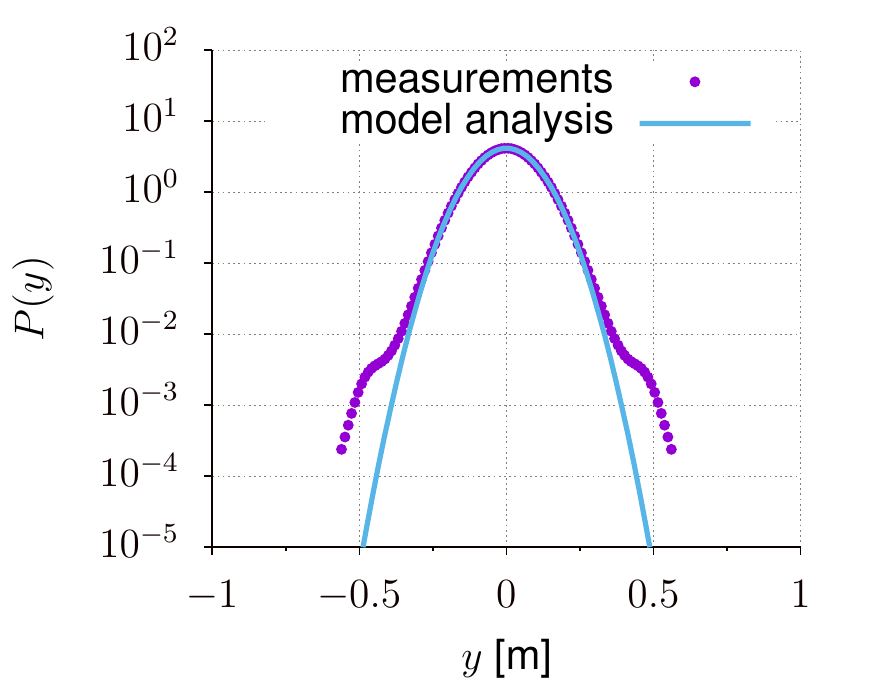}
};
\node at (0,-4.5){
       \includegraphics[width=5.8cm]{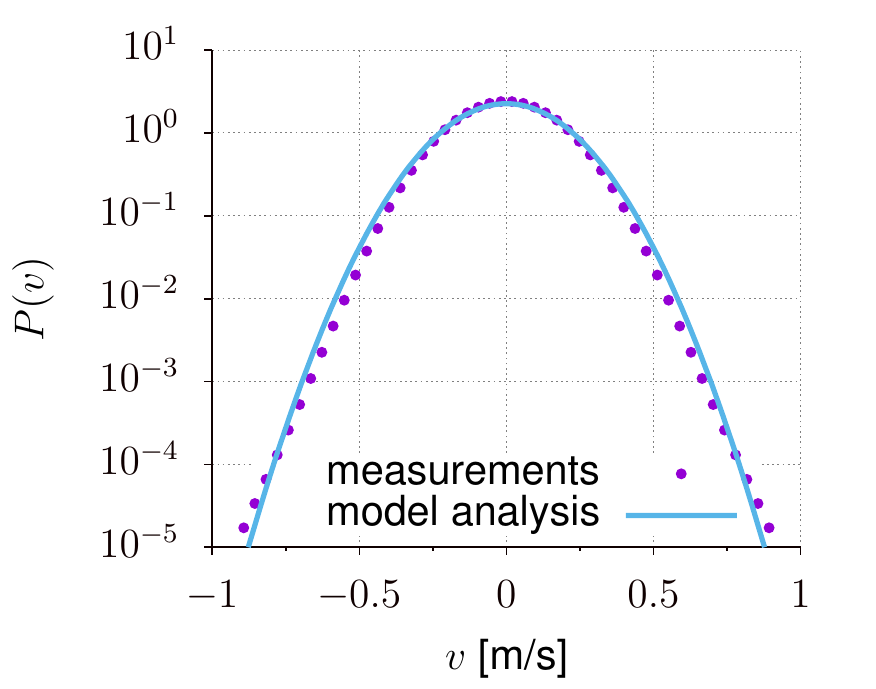}
};
\end{tikzpicture}
\caption{Transversal dynamics: comparison between measurements and model. We model the transversal motion as a harmonically bounded Langevin motion (cf. $y$ and $v$ dynamics in \eqref{m3} -- \eqref{m4}). In (a) we report the time-correlation function  of the transversal displacement $y$. The analytic solution (proportional to $\exp\left[-\gamma t\right]\left(\cos\omega t + \tfrac{\gamma}{\omega}\sin\omega t\right))$, with $\omega=\sqrt{2\beta -\gamma^2}$, see e.g.~\cite{risken1984fokker}) is reported as a blue  line. Measurements (colored dots) and simulations (empty dots) in a domain of equal size are in good agreement with simulations of the analytic solution. (b,c) Probability distribution function of, respectively, transversal positions $y$ and transversal velocities $v$. In both cases the analytic solution is a Gaussian distribution (blue line) which is in good agreement with the measurements (colored dots). In the case of transversal positions $y$ we observe rare deviations from the Gaussian behavior at $|y| > 0.4$. These are due to stopping events (cf. peak at $u = 0$ in Figure~\ref{figpotenvx}(c)).  We refer the reader to~\cite{corbetta2016fluctuations}  for further details on the calculations. Simulations included the same number of trajectories $N$ as in the observations ($N=72376$). (Figures reproduced from~\cite{corbetta2016fluctuations}).}
\label{figycorr}
\end{figure}

In the narrow corridor sketched in Figure~\ref{fig:corridor}, pedestrian entering on one side, for instance the left side, have just two options: either they reach the opposite right side walking with a velocity approximately equal to $u_m$ or, amid the corridor, they invert their direction and leave from the same (left) side, from which they (previously) entered. This last case corresponds to a transition of the walking velocity $u_m \rightarrow - u_m$. As the corridor has no source of interaction or distraction (walls are painted in white, there are no poster and no screens), we expect these transitions to be rather rare and  connected to external influencing factors: a change of thoughts, a phone call, etc. For the quasi 1D geometry, we expect no significant transversal dynamics in the corridor, beside small (Gaussian) oscillations.

For the sake of readability, and in view of the next analyses, we report here the model written in full and component-by-component. On this basis, we will  write the action $S[\gamma]$ explicitly and calculate its stationary points and the occurrence probability of rare events. For the individual position $\vec{x} = (x,y)$, we  identify, for convenience, with $x$ the longitudinal coordinate along the corridor and with $y$ the transversal coordinate. Similarly, for the velocity $\vec{v} = (u,v)$, we call $u$ and $v$ the velocity components in the $x$ and $y$ directions. Our modeling choice entails possibly the simplest dynamics encompassing: two stable velocity states, $\pm\vec{v}_m = (\pm u_m,0)$, motion confinement in the transversal direction, and no coupling between the two motion directions. Written in components, our model reads:
\begin{eqnarray}
\label{m1}
\dot{x}(t) &=& u(t) \\
\label{m2}
\dot u(t) &=& - \frac{\partial \phi(u,v)}{\partial u} + \sigma \dot{\eta}_x \\
\label{m3}
\dot{y}(t) &=& v(t) \\
\label{m4}
\dot{v}(t) &=& - \frac{\partial \phi(u,v)}{\partial v} - \frac{\partial V(v)}{\partial v} +  \sigma \dot{\eta}_y
\end{eqnarray}
where our modeling choice are specifically encoded in the potentials $\phi$ and $V$ that satisfy
\begin{eqnarray}
&&\phi(u,v) =  \alpha (u^2-u_m^2)^2 + \gamma v^2\\
&&V(y) = \beta y^2.
\end{eqnarray}
where $\alpha$, $\beta$ and $\gamma$ are positive model parameters. As usual, $\dot{\eta}_x$ and $\dot{\eta}_y$ are white delta-correlated -and mutually uncorrelated- Gaussian noises (with unit variance; cf. Table for numeric values adopted).

We model the transversal dynamics as a simple   damped linear harmonic oscillator with stochastic forcing. Three parameters regulate the transversal dynamics: $\gamma$, $\beta$ and $\sigma$. In Figure~\ref{figycorr}, we compare model and measurements in terms of three independent statistical properties: the time-autocorrelation function of the $y$ motion, and probability distribution of the $y$ and $v$ variables, i.e. the transversal position and the transversal velocity. Through these three quantities we can fix independently the values for the three parameters. We  find a good agreement between the probability distribution functions and the correlation functions measured and those produced by the model. This holds despite the simplicity of the model for the transversal dynamics, and provides an a posteriori justification of it.

\begin{figure*}
\centering
\begin{tikzpicture} 
\node at (0,0){
     \includegraphics[width=5.8cm]{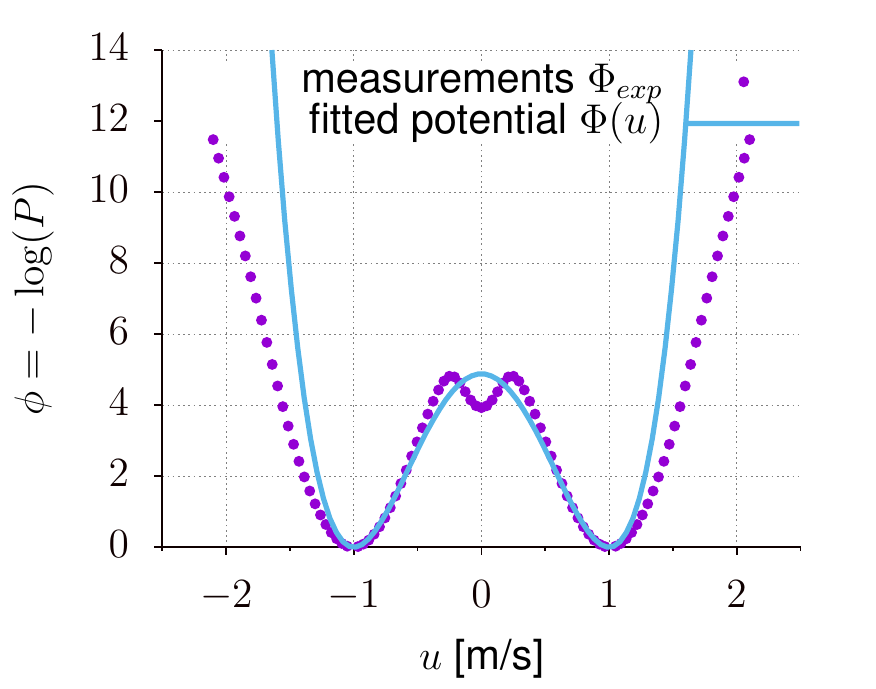}
     \includegraphics[width=5.8cm]{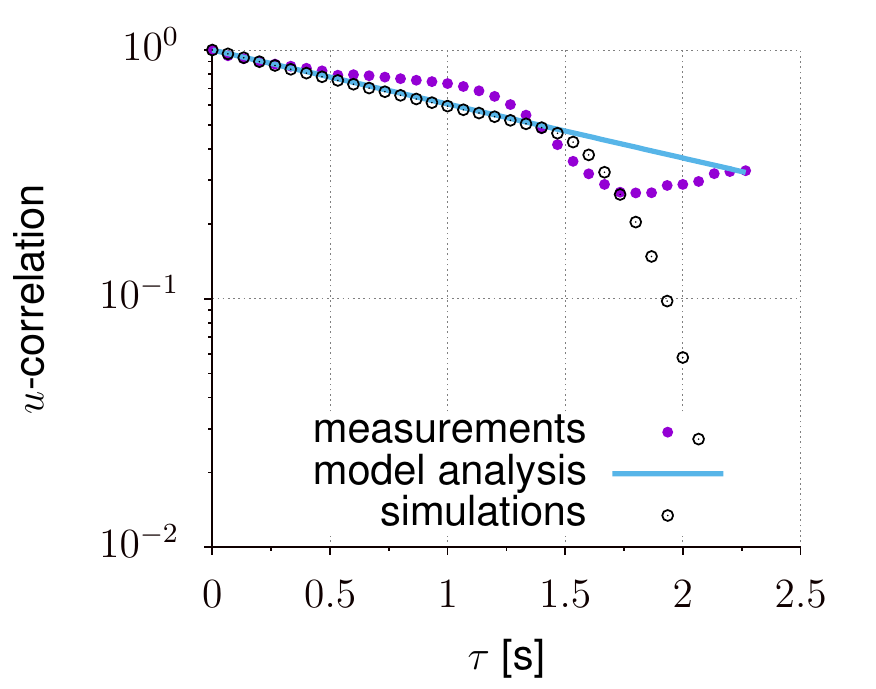}
};
\end{tikzpicture}
\caption{Longitudinal dynamics: comparison between measurements and model. We model the longitudinal motion as a Langevin dynamics in a double well velocity potential (cf. $x$ and $u$ dynamics in \eqref{m1}-\eqref{m2}).
In (a) we compare the potential obtained from field measurements (Eq.~\eqref{eq:expPot}, colored dots) with the rescaled potential $2/\sigma^2 \phi(u,0) = R (u^2 - u^2_p)^2$ (blue line). (b) Time correlation of the longitudinal velocity $u$. The analytic exponential decay of the linearized dynamics ($\exp(-8\alpha u_m^2t)$,  blue line) is compared with measurements (colored dots) and simulations of \eqref{m1} -- \eqref{m2} (in a simulated corridor with dimensions similar to those of our observations; empty dots). The finite size of the corridor is responsible for a deviation from an exponential decay: from simulations, we expect the correlation to decay exponentially  for small times only ($\tau < 1.5\,$s). The measured time correlation (cf.~\cite{corbetta2016fluctuations}  for detailed formulas) decays around the expected exponential trend  with larger discrepancies after $\tau > 0.75\,$s. Following the exponential decay at small times we fit the correlation time ($(8\alpha u_m^2)^{-1}$).  (Figures reproduced from~\cite{corbetta2016fluctuations}).} 
\label{figpotenvx}
\end{figure*}

The longitudinal dynamics is given by the simplest polynomial potential having minima providing stable velocities at $\pm u_m$. 
This choice notably encompasses also an unstable velocity state at $u=0$. 
As for the transversal velocity case, we can calibrate the parameters $\alpha$ and $\sigma$ of the model employing our measurements. The stationary probability distribution associated to Eq.~\eqref{m2} is $\Prob(u) = \frac{1}{M}\exp(-\frac{2}{\sigma^2}\phi(u,0))$ (cf. e.g.~\cite{risken1984fokker}), this enables, once  compared  with the measurements, to estimate the ratio $\frac{2\alpha}{\sigma^2}$. In particular, we fit the rescaled potential $\frac{2}{\sigma^2}\phi(u,0)$ to the (symmetric)  experimental potential
\begin{equation}\label{eq:expPot}
\Phi_{\exp}(u) = - \log \left(  \frac{\Prob_{\exp}(u) + \Prob_{\exp}(-u) }{2} \right),
\end{equation}
where $\Prob_{\exp}(u)$ is the probability distribution of $u$ observed experimentally. Specifically we fit to  recover the height of the potential well $\phi(0,0) - \phi(u_m,0)$ (cf. Figure~\ref{figpotenvx}(left)). Such a fit comes with two drawbacks: first, a poor agreement at high velocity; second, the approximation of the two unstable states at $u=\pm 0.2\,$m/s and of the stable state at $u=0$, through  the sole unstable state at $u=0$. This means that our model will underestimate the yet slight probability of remaining in $u=0$. 

To estimate $\alpha$ and $\sigma$ we need one further independent comparison with the data. We use the autocorrelation function of $u$. Note that we can have an analytic approximation of the time-correlation function of $u$ by linearizing Eq.~\eqref{m2} in the neighborhood of $u=u_m$. This yields a time correlation decaying as $\exp(-8\alpha u_m^2t)$ and thus the characteristic correlation time $(8\alpha u_m^2)^{-1}$, which provides a further relation on the parameters $\alpha$ and $\sigma$ that can now be fitted (cf. Figure~\ref{figpotenvx}(right)). 

\begin{figure}
\begin{center}
\centerline{\includegraphics[width=5.8cm]{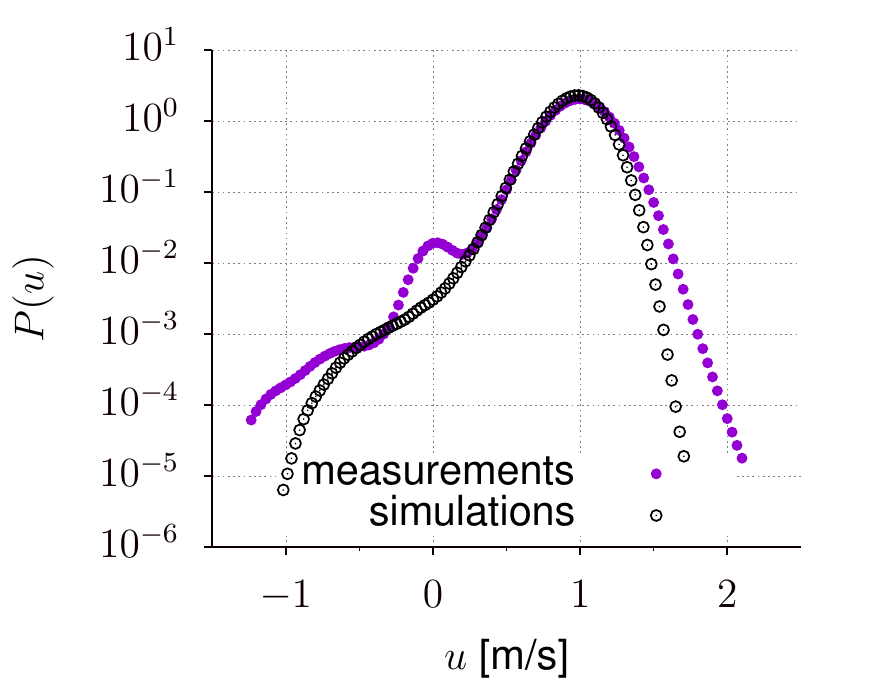} \includegraphics[width=5.8cm]{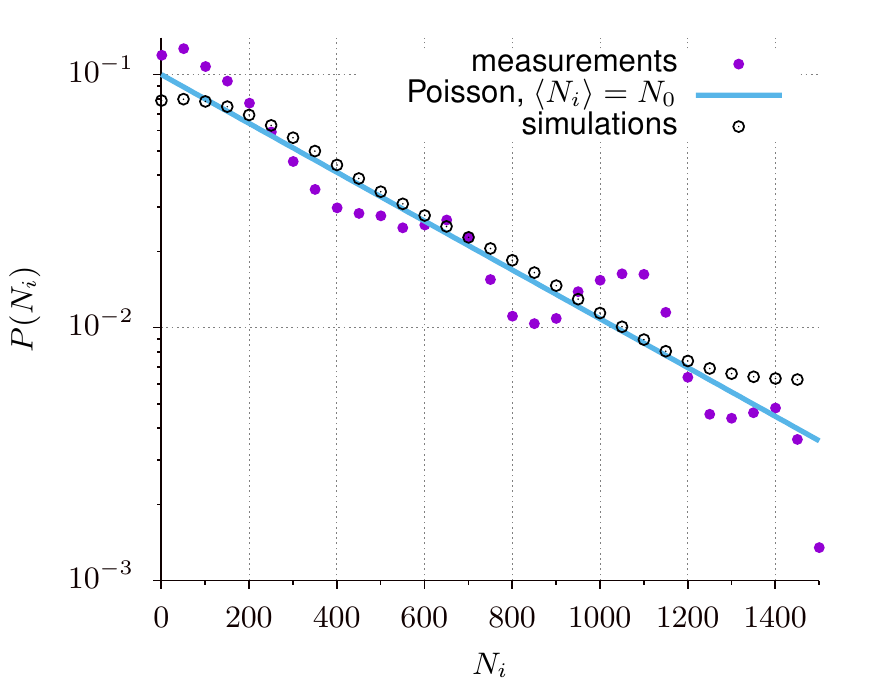}}
\caption{(a) Probability distribution function of longitudinal velocity $u$: comparison between  measurements (colored dots) and model (empty dots). The simulated dynamics
captures the entity of the fluctuation as well as the negative velocity tail within the
considered approximation (neglected high velocity behavior and stops).
(b)Probability distribution function of the number of pedestrians, $N_i$, passing in the corridor  between two trajectory inversion events (i.e. the number of consecutive crossings of the corridor). Comparison  of measurements (colored dots), simulation data from~\eqref{m1}-\eqref{m2} (black open circles) and of a Poisson process with expectation $\langle N_i\rangle =N_0 =  450$ pedestrians (blue line).  (Figures reproduced from~\cite{corbetta2016fluctuations}).}
\label{figturnback}
\end{center}
\end{figure}

With such an estimate of the parameters the model is  able to reproduce with good agreement not only the probability distribution function of the walking velocity $\Prob(u)$ (cf. Figure~\ref{figturnback}(left)) but also the statistics with which rare inversion events occur. In Figure~\ref{figturnback}(right) we report the Poisson distribution of inversion events in terms of the number, $N_i$, of pedestrians that we need to observe between two successive inversion events. 

We remark that the noise amplitude, $\sigma$, is here estimated twice and independently: once for the transversal and once for the longitudinal dynamics. These two estimates produced values in very strong agreement, which we retain as a consistency check for our modeling. In other words, our hypothesis of isotropic noise (a unique $\sigma$ constant appears in Eq.~\eqref{m2}-\eqref{m4}) is justified a posteriori. A further consistency check comes from the longitudinal and transversal correlation times which are extremely close (longitudinal correlation time: $1/(8\alpha u_m^2) \approx 2\,$s; transversal correlation time: $1/(2\gamma)\approx 2.4\,$s).

In the next section we use the path-integral representation to estimate the probability of the rare inversion events and connect them with the characteristic inversion time.

\subsection{Path-integral for the longitudinal dynamics}\label{sect:pathint-long}
Let us focus on the bi-stable longitudinal dynamics. The longitudinal velocity $u$ satisfies a first order stochastic ordinary differential equation (with, in our condition, no explicit coupling with the spatial variable $x$).

For the longitudinal velocity dynamics we can write the corresponding path-integral formulation, for which the Lagrangian~\eqref{eq:lagrangian-def} reads
\begin{equation}
\Lg(\dot u , u) = \frac{1}{2\sigma^2}\left(\dot u + \frac{\partial \phi}{\partial u} \right)^2.
\end{equation}
Hence, the stationary trajectories for the action are the solutions of the Euler-Lagrange equation
\begin{equation}
\frac{\partial \Lg}{\partial u} - \frac{d}{dt}\frac{\partial \Lg}{\partial \dot u} = 0,
\end{equation}
which satisfy 
\begin{equation}
\ddot u = -\frac{1}{2}\frac{\partial }{\partial u}\left(\frac{\partial \phi}{\partial u}\right)^2
\end{equation}
i.e. trajectories follow the Hamiltonian field generated by the Hamiltonian function
\begin{equation}
H(\dot u, u) = \frac{1}{2}\dot u^2 + \frac{1}{2}\left(\frac{\partial \phi}{\partial u}\right)^2.
\end{equation}
The solutions satisfy
\begin{equation}\label{eq-stationary-u}
\dot u = \pm \frac{\partial \phi}{\partial u}.
\end{equation}
We stress  that, so far, no assumption has been done on the structure of $\phi$.

The solutions of~\eqref{eq-stationary-u} relate  with the solutions of~\eqref{m2} for vanishing noise ($\sigma = 0$). In particular, one of the two solutions coincide with the case of vanishing noise ($\dot u = - \frac{\partial \phi}{\partial u}$) while the other involves also the inversion of the sign of the potential ($\dot u = + \frac{\partial \phi}{\partial u}$). Remarkably, the first set of motions entails the descent to the bottom of the potential wells to the global minimum $u=u_m$, while the second set entails  the ascent toward the local maximum of the potential at $u=0$. Once more, these dynamics are the local extrema for the probability density $\rho[\gamma]$. 

Let us compare the occurrence probability of these two opposite behavior when it comes to descend or ascend the potential $\phi$. 

Reaching the bottom of the potential well ($\dot u = - \frac{\partial \phi}{\partial u}$) involves energy dissipation, thus we expect relatively high occurrence probability. In this case $\rho[\gamma_{0\rightarrow u_m}]$, for a motion $\gamma$ between $u=0$ and $u=u_m$ reads
\begin{equation}
\rho[\gamma_{0\rightarrow u_m}] = \frac{1}{M}e^{\frac{1}{2\sigma^2}\int_{u=0}^{u=u_m} dt \left(\dot u - \frac{\partial \phi}{\partial u}\right)^2} = \frac{1}{M},
\end{equation}
where the last equality follows the fact that the integrand function is identically zero. 
On the contrary, if we ascend the potential well ($\dot u =  \frac{\partial \phi}{\partial u}$) from $u=u_m$ to $u=0$ (trajectory $\gamma_{u_m\rightarrow 0}$) it holds
\begin{equation}
\rho[\gamma_{u_m\rightarrow 0}] = \frac{1}{M}e^{-\frac{1}{2\sigma^2}\int_{u=u_m}^{u=0}dt \left( \frac{\partial \phi}{\partial u} + \frac{\partial \phi}{\partial u}\right)^2  } =  \frac{1}{M}e^{-\frac{2}{\sigma^2}\int_{u=u_m}^{u=0}dt \left(\frac{\partial \phi}{\partial u}\right)^2  }.
\end{equation}
We can evaluate the integral at the exponent as follows
\begin{align}
\int_{u=u_m}^{u=0}dt  \left( \frac{\partial \phi}{\partial u}\right)^2 = \int_{u=u_m}^{u=0}dt   \dot u \frac{\partial \phi}{\partial u} = \int_{u_m}^{0} du    \frac{\partial \phi}{\partial u} = \phi(0) - \phi(u_m).
\end{align}
Therefore, we obtain
\begin{equation}
\rho[\gamma_{u_m\rightarrow 0}] = \frac{1}{M}e^{- \frac{2}{\sigma^2} \phi(0) - \phi(u_m)}.
\end{equation}
The ratio 
\begin{equation}\label{eq:kramer-est}
\frac{\rho[\gamma_{u_m\rightarrow 0}]}{\rho[\gamma_{0\rightarrow u_m}]} = e^{- \frac{2}{\sigma^2} \phi(0) - \phi(u_m)} = e^{- \frac{2\alpha}{\sigma^2}u_m^4}
\end{equation}
which compares the probability of the rare ascents, to the common descents, corresponds to the well-known Kramer's estimate for the probability of rare inversion events~\cite{kramers1940brownian}. Note that the probability of rare events gets exponentially smaller as the term $R = 2(\phi(0) - \phi(u_m))/\sigma^2$ increases, i.e. when the potential barrier  $\phi(0) - \phi(u_m)$ is larger or the noise intensity diminishes. Non surprisingly, the ratio in Eq.~\eqref{eq:kramer-est} further gives the scale of  the characteristic time of inversion events $T_i$, i.e. the average time necessary to escape from the bottom of the well and reach the unstable state $u=0$. Time dimensions are given by a prefactor (cf. e.g.~\cite{benzi1981mechanism})  as  
\begin{equation}
T_i  \approx \frac{\pi}{\sqrt{  \frac{\partial^2\phi}{\partial^2 u} |_{u=u_m}  - \left|\frac{\partial^2\phi}{\partial^2 u}|_{u=0} \right| }  } e^{- \frac{2\alpha}{\sigma^2}u_m^4}.
\end{equation}
From this estimate, the number pedestrians that we need to observe between two inversion events $N_i$ is
\begin{equation}
N_i = T_i/T_c
\end{equation}
where $T_c$ is the characteristic time for crossing (in our case $T_c\approx 1.8\,$s). The obtained value of $N_i$ remains in good agreement with our experimental observations (we obtain $T_i\approx 1000\,$s, which yields $N_i\approx 555$; cf. Figure~\ref{figturnback}(right)).

The path-integral formulation gives furthermore a detailed insight in the dynamics that brings to a velocity inversion. Velocity inversion events have average trajectory $\dot u = + \frac{\partial \phi}{\partial u}$. This trajectory entails a gradual ascent of the velocity potential well up to its top. In consideration of the dynamics~\eqref{m2}, this can only happen with a sequence of ``favorable'' outcomes of the random forcing that are opposite (and double in intensity) of the friction-like descent force  $-\frac{\partial \phi}{\partial u}$. In other words, inversion events are not, e.g., outcomes of an  impulsive event  in the direction opposite to the motion. Rather, they occur gradually, as a chain of small solicitations, that ultimately bring to an inversion of the direction of motion.

\section{Discussion}\label{sect:discussion}
In this chapter we discussed the usage of the path-integration formalism as a modeling framework for pedestrian dynamics. Path-integration provides a trajectory-centric modeling tool assigning to each physical pedestrian trajectory the probability that it is observed. Considering the complexity of pedestrian dynamics in real-life venues, where pedestrian trajectories distribute among different usage patterns, for each of which they show large variabilities, a tool focusing on the observational probabilities of trajectories seems a most natural and intuitive representation choice.

In the path-integral conceptual framework, the knowledge of the system is given by the action functional, $\Sf[\gamma]$, which represents the theory and encodes for all available knowledge, e.g.  $\Sf[\gamma]$ allows to  fully characterize the statistical behavior, and the  usage patterns including rare events.  In this chapter, we wrote a quantitatively accurate action functional for the case of the diluted pedestrian dynamics in a narrow corridor. The description we gave is equivalent to the Langevin model that we obtained in our previous work~\cite{corbetta2016fluctuations}, yet for its direct connection with trajectories it is suited for generalizations. Through the action functional we could furthermore recover the behaviors that are local extrema of the observation probability and estimate the probability of rare events (U-turns).

We focused on the diluted limit of pedestrian dynamics, for which pedestrian-pedestrian interactions remain negligible. As in the original quantum mechanical formulation, extension involving multiple pedestrians up to dense dynamics are possible.

\section*{Acknowledgments}
The thank Roberto Benzi (Rome, IT) for  useful discussions. We acknowledge the support of Naturalis Biodiversity Center for hosting our measurement setup.  This work is part of the JSTP research programme ``Vision driven visitor behaviour analysis and crowd management'' with project number 341-10-001, which is financed by the\
 Netherlands Organisation for Scientific Research (NWO).

\bibliography{master}

\begin{thebibliography}{10}

\bibitem{benzi1981mechanism}
R.~Benzi, A.~Sutera, and A.~Vulpiani.
\newblock The mechanism of stochastic resonance.
\newblock {\em Journal of Physics A}, 14(11):L453, 1981.

\bibitem{chow2015path}
C.~Chow and M.~Buice.
\newblock Path integral methods for stochastic differential equations.
\newblock {\em Journal of Mathematical Neuroscience}, 5(1):8, 2015.

\bibitem{corbetta2014TRP}
A.~Corbetta, L.~Bruno, A.~Muntean, and F.~Toschi.
\newblock High statistics measurements of pedestrian dynamics.
\newblock {\em Transportation Research Procedia}, 2:96--104, 2014.

\bibitem{corbetta2016fluctuations}
A.~Corbetta, C.~Lee, R.~Benzi, A.~Muntean, and F.~Toschi.
\newblock Fluctuations around mean walking behaviours in diluted pedestrian
  flows.
\newblock {\em Physical Review E}, 95:032316, 2017.

\bibitem{corbettaTGF15}
A.~Corbetta, C.~Lee, A.~Muntean, and F.~Toschi.
\newblock Asymmetric pedestrian dynamics on a staircase landing from continuous
  measurements.
\newblock In W.~Daamen and V.~Knoop, editors, {\em Traffic and Granular Flows
  '15}, chapter~7. Springer, 2016.

\bibitem{CDA10}
A.~Corbetta, C.~Lee, A.~Muntean, and F.~Toschi.
\newblock Frame vs. trajectory analyses of pedestrian dynamics asymmetries in a
  staircase landing.
\newblock {\em Collective Dynamics}, 1:1--26, 2017.

\bibitem{corbetta2016continuous}
A.~Corbetta, J.~Meeusen, C.~Lee, and F.~Toschi.
\newblock Continuous measurements of real-life bidirectional pedestrian flows
  on a wide walkway.
\newblock In {\em Pedestrian and Evacuation Dynamics 2016}, pages 18--24.
  University of Science and Technology of China press, 2016.

\bibitem{corbetta2015parameter}
A.~Corbetta, A.~Muntean, and K.~Vafayi.
\newblock Parameter estimation of social forces in pedestrian dynamics models
  via a probabilistic method.
\newblock {\em Mathematical Biosciences and Engineering}, 12(2):337--356, 2015.

\bibitem{datasetped}
A.~Corbetta and F.~Toschi.
\newblock Crowdflow – diluted pedestrian dynamics in the metaforum building
  of eindhoven university of technology, 2017.

\bibitem{cristiani2014multiscale}
E.~Cristiani, B.~Piccoli, and A.~Tosin.
\newblock {\em Multiscale Modeling of Pedestrian Dynamics}, volume~12.
\newblock Springer, 2014.

\bibitem{durr1978onsager}
D.~D{\"u}rr and A.~Bach.
\newblock The onsager-machlup function as lagrangian for the most probable path
  of a diffusion process.
\newblock {\em Communications in Mathematical Physics}, 60(2):153--170, 1978.

\bibitem{ester1996density}
M.~Ester, H.~Kriegel, J.~Sander, X.~Xu, et~al.
\newblock A density-based algorithm for discovering clusters in large spatial
  databases with noise.
\newblock In {\em Proceedings of the Second International Conference on
  Knowledge Discovery and Data Mining, Kdd-96}, pages 226--231, 1996.

\bibitem{helbing2001traffic}
D.~Helbing.
\newblock Traffic and related self-driven many-particle systems.
\newblock {\em Reviews of modern physics}, 73(4):1067, 2001.

\bibitem{helbing1995social}
D.~Helbing and P.~Moln{\'a}r.
\newblock Social force model for pedestrian dynamics.
\newblock {\em Physical Review E}, 51(5):4282, 1995.

\bibitem{helbing2001self}
D.~Helbing, P.~Moln{\'a}r, I.~Farkas, and K.~Bolay.
\newblock Self-organizing pedestrian movement.
\newblock {\em Environment and Planning B}, 28(3):361--383, 2001.

\bibitem{hoogendoorn2007microscopic}
S.~Hoogendoorn and W.~Daamen.
\newblock Microscopic calibration and validation of pedestrian models:
  Cross-comparison of models using experimental data.
\newblock In {\em Traffic and Granular Flow '05}, pages 329--340. Springer,
  2007.

\bibitem{klebaner2012introduction}
F.~Klebaner.
\newblock {\em Introduction to stochastic calculus with applications}.
\newblock World Scientific Publishing Company, 2012.

\bibitem{kramers1940brownian}
H.~A. Kramers.
\newblock Brownian motion in a field of force and the diffusion model of
  chemical reactions.
\newblock {\em Physica}, 7(4):284--304, 1940.

\bibitem{kwak2013collective}
J.~Kwak, H.~Jo, T.~Luttinen, and I.~Kosonen.
\newblock Collective dynamics of pedestrians interacting with attractions.
\newblock {\em Physical Review E}, 88(6):062810, 2013.

\bibitem{lemons1997paul}
D.~Lemons and A.~Gythiel.
\newblock Paul langevin's 1908 paper ``{O}n the {T}heory of {B}rownian
  {M}otion'' (``{S}ur la th{\'e}orie du mouvement brownien'' cr acad.
  sci.(paris) 146, 530-533 (1908)).
\newblock {\em American Journal of Physics}, 65:1079--1081, 1997.

\bibitem{Kinect}
{Microsoft Corp.}
\newblock Kinect for {X}box 360, available online:
  http://www.xbox.com/en-us/kinect/, 2011.
\newblock Redmond, WA, USA.

\bibitem{risken1984fokker}
H.~Risken.
\newblock {\em Fokker-Planck Equation}.
\newblock Springer, Berlin, 1984.

\bibitem{Lutz}
P.~Romanczuk, M.~B{\"a}r, W.~Ebeling, B.~Lindner, and L.~Schimansky-Geier.
\newblock Active {B}rownian particles.
\newblock {\em The European Physical Journal Special Topics}, 202(1):1--162,
  2012.

\bibitem{schadschneider2001cellular}
A.~Schadschneider.
\newblock Cellular automaton approach to pedestrian dynamics-theory.
\newblock {\em arXiv preprint cond-mat/0112117}, 2001.

\bibitem{seer2014kinects}
S.~Seer, N.~Br{\"a}ndle, and C.~Ratti.
\newblock Kinects and human kinetics: A new approach for studying pedestrian
  behavior.
\newblock {\em Transportation Research C}, 48:212--228, 2014.

\bibitem{seyfried2009new}
A.~Seyfried, O.~Passon, B.~Steffen, M.~Boltes, T.~Rupprecht, and W.~Klingsch.
\newblock New insights into pedestrian flow through bottlenecks.
\newblock {\em Transportation Science}, 43(3):395--406, 2009.

\bibitem{zinn1996quantum}
J.~Zinn-Justin.
\newblock {\em Quantum field theory and critical phenomena}.
\newblock Clarendon Press, 1996.

\end{thebibliography}

\end{document}